\documentclass[aip, reprint, floatfix,twocolumn]{revtex4-1}
\usepackage{graphicx}
\usepackage{wrapfig}
\usepackage{dcolumn}
\usepackage{bm}
\usepackage{here}
\usepackage{amsfonts} 
\usepackage{latexsym}
\begin{document}

\title{
Equation-of-motion approach of spin-motive force
}

\author{Yuta Yamane$^{1,2}$}
\author{Jun'ichi Ieda$^{1,3}$}
\author{Jun-ichiro Ohe$^{1,3}$}
\author{Stewart E. Barnes$^{4}$}
\author{Sadamichi Maekawa$^{1,3}$}
\affiliation{
$^{1}$Advanced Science Research Center, Japan Atomic Energy Agency, Tokai, Ibaraki 319-1195, Japan \\
$^{2}$Institute for Materials Research, Tohoku University, Sendai 980-8577, Japan \\
$^{3}$CREST, Japan Science and Technology Agency, Tokyo 102-0075, Japan\\
$^{4}$Physics Department, University of Miami, Coral Gables, FL33124, USA
}

 \date{\today}

\begin{abstract}

We formulate a quantitative theory of an electromotive force of spin origin, i.e., spin-motive force, by the equation-of-motion approach.
In a ferromagnetic metal, electrons couple to the local magnetization via the exchange interaction.
The electrons are affected by spin dependent forces due to this interaction and the spin-motive force and the anomalous Hall effect appears.
We have revealed that the origin of these phenomena is a misalignment between the conduction electron spin and the local magnetization.

\end{abstract}

\maketitle

%
%

\section*{introduction}

The study of the spin dependent transport properties of ferromagnets is one of the most importrant subjects from both a fundamental and an application points of view.
In a ferromagnetic metal, the spin of a conduction electron interacts with the local magnetization through the exchange coupling, and this interaction strongly affects the electron charge dynamics.
Recently, an electromotive force of spin origin which is called a spin-motive force, has been theoretically predicted\cite{barnes} and experimentally observed\cite{yang,exp2}.
The spin-motive force reflects the conversion from the magnetic energy of a ferromagnet to the electric energy of electrons.
The spin-motive force is essentially due to the exchange interaction between the electron spin and the local magnetization and provides a new concept of electric devices.

In this paper, we show that the origin of the spin-motive force is a misalignment between the electron spin and the local magnetization by using the equation-of-motion approach.
We investigate a force acting upon an electron due to the exchange interaction, which accelerates electrons in the direction in which the exchange energy decreases. 
We derive the formula of the force, which is spin dependent.
The spin dependent force consists of a driving force and a Lorentz type force which contribute to the spin-motive force and the anomalous Hall effect \cite{ahe1,ahe2,ahe3,ahe4,ahe5}, respectively.

The exsistence of such spin dependent forces has been argued in the past studies \cite{barnes,yang,berger,volovik,stern,stone,saslow,duine,sanvito,tserkov}.
However, our approach provides a more intutitive and physical understanding of the origin of the spin-motive force; 
if a misalignment arises between the electron spin and the local magnetization, spatial inhomogenity of the electric potential energy appears due to the spatial inhomogenity of the exchange energy.

%
%

\section*{model calculation}

As disccused above, in a ferromagnet the electron spin couples to the magnetization via the exchange interaction. 
We treat this interaction in the s-d Hamiltonian given by
\begin{equation}
\mathcal{H}=\frac{\textrm{\boldmath $p$}^2}{2m}-J\textrm{\boldmath $\sigma$}\cdot\textrm{\boldmath $m$}\left(x,t\right),
\label{hamiltonian}
\end{equation}
where $\textrm{\boldmath $p$}$, $m$ ,$J$ and $\textrm{\boldmath $\sigma$}$ are the momentum operator, the electron mass, the exchange coupling energy and the Pauli's matrices, respectively. 
the unit vector of the magnetization direction is represented by $\textrm{\boldmath $m$}$ which has spatial and time dependence.

\begin{figure}[t]
  \centerline{\includegraphics[width=50mm]{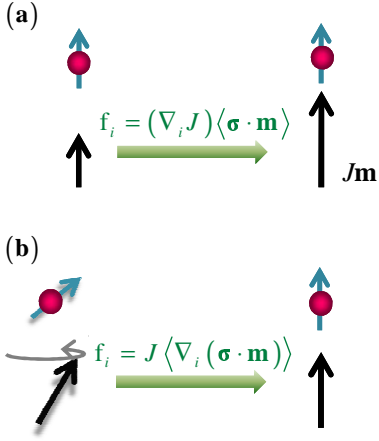}}
  \caption{
   Forces acting upon an electron spin $\textrm{\boldmath $\sigma$}$ due to the exchange coupling $J$ with the local magnetization $\textrm{\boldmath $m$}$.
 }
 \label{fig:fig1}
\end{figure}

In order to examine the electron dynamics, we consider the equation of motion of the electron which is obtained by the Heisenberg equation as
\begin{eqnarray}
 f_i &=& m\partial_t^2x_i \nonumber\\
     &=& \frac{m}{\left(i\hbar\right)^2}\left[\left[x_i,\mathcal{H}\right],\mathcal{H}\right]\nonumber\\
     &=& \nabla_i\left(J\textrm{\boldmath $\sigma$}\cdot\textrm{\boldmath $m$}\right) \label{gradient}\nonumber\\
     &=& \left(\nabla_iJ\right)\textrm{\boldmath $\sigma$}\cdot\textrm{\boldmath $m$}+J\nabla_i\left(\textrm{\boldmath $\sigma$}\cdot\textrm{\boldmath $m$}\right),
\label{force1}
\end{eqnarray}
where $x_i$ and $f_i$ are the $i$-components ($i=x,y,z$) of the position and ``force'' operators for the electron, respectively.
The ``force'' is given by the spatial derivative of the exchange energy; 
the electron is accelerated in the direction in which the potential energy $-J\textrm{\boldmath $\sigma$}\cdot\textrm{\boldmath $m$}$ decreases. 
The first term in Eq.~(\ref{force1}) reflects the spatial inhomogenity of the exchange energy $J$ [Fig.~\ref{fig:fig1}(a)],
while the second term comes from the spatial inhomogenity of $\textrm{\boldmath $\sigma$}\cdot\textrm{\boldmath $m$}$ [Fig.~\ref{fig:fig1}(b)].
Hereafter, we consider the case in which the $J$ is constant and focus on the effects of the second term.
The net force acting upon the electron is
\begin{equation}
\left\langle f_i \right\rangle = J \left\langle \textrm{\boldmath $\sigma$} \right\rangle \cdot \nabla_i\textrm{\boldmath $m$}.
\label{force2}\end{equation}
The expected value $\left\langle\cdots\right\rangle$ means $\mathrm{Tr}(\rho\cdots)$, $\rho$ is the $2\times2$ density matrix in the spin space, and $\left\langle \textrm{\boldmath $\sigma$} \right\rangle$ is the unit vector of the electron spin direction.

To calculate the force (\ref{force2}), we have to determine the dynamics of the magnetization $\textrm{\boldmath $m$}$ and the electron spin $\left\langle\textrm{\boldmath $\sigma$}\right\rangle$. 
It is well known that the magnetization $\textrm{\boldmath $m$}$ obeys the phenomenological Landau-Lifshiz (LL) equation
\begin{equation}
\partial_t\textrm{\boldmath $m$}=-\gamma\textrm{\boldmath $m$}\times\textrm{\boldmath $H$}-\lambda\textrm{\boldmath $m$}\times\left(\textrm{\boldmath $m$}\times\textrm{\boldmath $H$}\right),
\label{ll}\end{equation}
where $\gamma$ is the gyromagnetic ratio, $\lambda$ the damping constant, and $\textrm{\boldmath $H$}=-(1/M_S\mu_0)\delta F/\delta\textrm{\boldmath $m$}$ is the effective magnetic field which includes the external field, the exchange field due to the ferromagnetic coupling, and any other effects that are described in the free energy $F[M_S\textrm{\boldmath $m$}]$.
$M_S$ and $\mu_0$ are the saturation magnetization and the magnetic constant, respectively.
Once the explicit form of the free energy $F[M_S\textrm{\boldmath $m$}]$ is given, the magnetization dynamics $\textrm{\boldmath $m$}$ can be obtained by solving Eq.~(\ref{ll}).

The remaining work is to determine the electron spin dynamics $\left\langle\textrm{\boldmath $\sigma$} \right\rangle$, which prefers to be parallel to the magnetization $\textrm{\boldmath $m$}$ due to the strong exchange interaction between them. 
Before concrete disccusions, let us make a point. 
In general, the electron spin $\left\langle\textrm{\boldmath $\sigma$} \right\rangle$ is expressed as $\left\langle\textrm{\boldmath $\sigma$} \right\rangle=C\textrm{\boldmath $m$}+D\delta\textrm{\boldmath $m$}$, where $\delta\textrm{\boldmath $m$}$ is a unit vector perpendicular to $\textrm{\boldmath $m$}$, and $C$ and $D$ are coefficients ($\left| C \right|\gg \left| D \right|$). 
The component $C\textrm{\boldmath $m$}$ does not contribute to the force (\ref{force2}), because $\textrm{\boldmath $m$}\cdot\nabla_i\textrm{\boldmath $m$}=0$.
In ohter words, if the electron spin is exactly parallel to the magnetization at every time and space, the electron's energy is the same always and everywhere, and the electron does not feel any force.
This means that what causes the force is the off-diagonal component of the electron spin with respect to the magnetization, i.e., the misalignment between them.

According to the Heisenberg equation, the equation of motion of the electron spin is 
\begin{equation}
\partial_t\left\langle \textrm{\boldmath $\sigma$}\right\rangle = \frac{J\hbar}{2}\left\langle\textrm{\boldmath $\sigma$}\right\rangle\times\textrm{\boldmath $m$}\left(x,t\right).
\label{larmor}\end{equation}
This expression shows the Larmor precession of the conduction electron spin around the magnetization.
Actually, in addition to the precession torque, the electron spin feels a damping torque to the magnetization.
To obtain an analytical form of the electron spin dynamics without solving Eq.~(\ref{larmor}) in a straightforward manner, let us consider the problem in a more convenient coordinate system in which the magnetization is parallel to the $z$-axis at every time and space.
Such an operation is realized by a SU(2) local gauge transformation $\psi\mapsto U\psi$, where $U=\exp\left(i\sigma_y\theta\left(x,t\right)/2\right)\exp\left(i\sigma_z\varphi\left(x,t\right)/2\right)$, 
$\theta$ and $\varphi$ are defined by $\textrm{\boldmath $m$}=\left(\sin\theta\cos\varphi,\sin\theta\sin\varphi,\cos\theta\right)$, and $\psi$ is a two component spinor.

After performing the gauge transformation, the Hamiltonian in the rotational frame is
\begin{eqnarray}
 \mathcal{H}' &=& U\mathcal{H}U^\dagger-i\hbar U\left(\partial_t U^\dagger\right)\nonumber\\
              &=& \frac{\left(\textrm{\boldmath $p$}-i\hbar U\left(\nabla U^\dagger\right)\right)^2}{2m}-J\sigma_z-i\hbar U\left(\partial_t U^\dagger\right)\nonumber\\
              &=& \frac{\left(\textrm{\boldmath $p$}-\sigma_x\textrm{\boldmath $A$}_x-\sigma_y\textrm{\boldmath $A$}_y-\sigma_z\textrm{\boldmath $A$}_z\right)^2}{2m}-J\sigma_z-i\hbar U\left(\partial_t U^\dagger\right)\nonumber\\
              &=& \frac{\textrm{\boldmath $p$}^2+\textrm{\boldmath $A$}_x^2+\textrm{\boldmath $A$}_y^2+\textrm{\boldmath $A$}_z^2}{2m}-\textrm{\boldmath $\sigma$}\cdot\left(J\textrm{\boldmath $m$}'+\textrm{\boldmath $d$}'\right),\nonumber\\
\label{hamiltonian2}
\end{eqnarray}
where
\begin{eqnarray}
\textrm{\boldmath $d$}'&=&\frac{\hbar}{2}\left(\textrm{\boldmath $m$}\times\partial_t\textrm{\boldmath $m$}\right)'\nonumber\\
&&+\frac{\hbar\cos\theta\partial_t\varphi}{2}\textrm{\boldmath $m$}'
+\frac{1}{2m}\left[
\begin{array}{c}
\textrm{\boldmath $p$}\cdot\textrm{\boldmath $A$}_x+\textrm{\boldmath $A$}_x\cdot\textrm{\boldmath $p$} \\
\textrm{\boldmath $p$}\cdot\textrm{\boldmath $A$}_y+\textrm{\boldmath $A$}_y\cdot\textrm{\boldmath $p$}  \\
\textrm{\boldmath $p$}\cdot\textrm{\boldmath $A$}_z+\textrm{\boldmath $A$}_z\cdot\textrm{\boldmath $p$}   
\end{array}\right]
.
\end{eqnarray}
The SU(2) pure gauge fields, $i\hbar U\left(\nabla U^\dagger\right)=\sigma_x\textrm{\boldmath $A$}_x+\sigma_y\textrm{\boldmath $A$}_y+\sigma_z\textrm{\boldmath $A$}_z$ and $-i\hbar U\left(\partial_t U^\dagger\right)$, arise as a result of the gauge transformation.
The superscript $'$ denotes a vector in the rotational frame, for instance, $\textrm{\boldmath $m$}'=(0,0,1)$.

\begin{figure}[t]
  \centerline{\includegraphics[width=20mm]{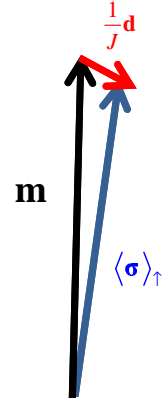}}
  \caption{
   The misalignment between the electron spin $\textrm{\boldmath $\sigma$}$ and the local magnetization $\textrm{\boldmath $m$}$.
 }
 \label{fig:fig2}
\end{figure}

Equation (\ref{hamiltonian2}) indicates that when the magnetization depends upon time, the electron spin feels the effective magnetic field $J\textrm{\boldmath $m$}'+\textrm{\boldmath $d$}'$ in the rotating frame.
Therefore, in reality, the electron spin precesses around the effective magnetic field $J\textrm{\boldmath $m$}'+\textrm{\boldmath $d$}'$ rather than the magnetization $J\textrm{\boldmath $m$}'$.
If the magnetic texture is smooth and slow enough, we can adopt an adiabatic approximation in which $\left\langle\textrm{\boldmath $\sigma$} \right\rangle$ is approximated to be parallel or antiparallel to the effective magnetic field at every time and space; 
a normalized electron spin is
\begin{eqnarray}
\left\langle\textrm{\boldmath $\sigma$} \right\rangle'_{\uparrow\downarrow}
&=&\pm\textrm{\boldmath $m$}'\pm\frac{\hbar}{2J}(\textrm{\boldmath $m$}\times\partial_t\textrm{\boldmath $m$})'\nonumber\\
&&\pm\frac{\hbar\cos\theta\partial_t\varphi}{2J}\textrm{\boldmath $m$}'
\pm\frac{1}{mJ}\left[
\begin{array}{c}
\textrm{\boldmath $A$}_x\cdot m\textrm{\boldmath $v$}  \\
\textrm{\boldmath $A$}_y\cdot m\textrm{\boldmath $v$}  \\
\textrm{\boldmath $A$}_z\cdot m\textrm{\boldmath $v$}  
\end{array}
\right],
\end{eqnarray}
where $\uparrow(\downarrow)$ stands for the majority (minority) spin with respect to the local magnetization and corresponds to $+$($-$) in the right-hand side.
Here, we replace $\textrm{\boldmath $p$}\cdot\textrm{\boldmath $A$}_i+\textrm{\boldmath $A$}_i\cdot\textrm{\boldmath $p$}$ by $2\textrm{\boldmath $A$}_i\cdot m\textrm{\boldmath $v$}$, where $\textrm{\boldmath $v$}$ is the averaged velocity vector of electrons.
The expression for $\left\langle\textrm{\boldmath $\sigma$} \right\rangle$ in the laboratory frame is obtained by the $O(3)$ rotation as
\begin{eqnarray}
\left\langle\textrm{\boldmath $\sigma$} \right\rangle _{\uparrow\downarrow}&=&\left[
\begin{array}{ccc}
\cos\theta\cos\varphi & -\sin\varphi & \sin\theta\cos\theta \\
\cos\theta\sin\varphi & \cos\varphi & \sin\theta\sin\varphi \\
-\sin\theta & 0 & \cos\theta
\end{array}
\right]\left\langle\textrm{\boldmath $\sigma$} \right\rangle _{\uparrow\downarrow}'\nonumber\\
&=& \pm \left(\textrm{\boldmath $m$}+\textrm{\boldmath $d$}/J\right).
\label{spin}\end{eqnarray}
That is, the electron spin has the off-diagonal component with respect to the magnetization, $\textrm{\boldmath $d$}/J$ (FIG.~\ref{fig:fig2}).

By substituting Eq.~(\ref{spin}) to Eq.~(\ref{force2}), the force can be expressed only by the magnetization dynamics as
\begin{equation}
\left\langle f_i \right\rangle _{\uparrow\downarrow}=\pm \frac{\hbar}{2}\left(\textrm{\boldmath $m$}\times\partial_t\textrm{\boldmath $m$}\right)\cdot\nabla_i\textrm{\boldmath $m$}-e\textrm{\boldmath $v$}\times\textrm{\boldmath $B$}_{\uparrow\downarrow},
\label{force3}\end{equation}
where
\begin{equation}
\left(\textrm{\boldmath $B$}_{\uparrow\downarrow}\right)_i=\mp\frac{\hbar}{4e}\epsilon^{ijk}\textrm{\boldmath $m$}\cdot\left(\nabla_j\textrm{\boldmath $m$}\times\nabla_k\textrm{\boldmath $m$}\right).
\label{magnetic}\end{equation}
The first term in Eq.~(\ref{force3}) is a spin dependent driving force, which contributes to the spin-motive force.
On the other hand, the second term in Eq.~(\ref{force3}) is the Lorentz type force due to the spin dependent magnetic field $\textrm{\boldmath $B$}_{\uparrow\downarrow}$, produced by the non-coplanar magnetization configuration as Eq.~(\ref{magnetic}), and the transverse conductivity appears, i.e., the anomalous Hall effect \cite{ahe1,ahe2,ahe3,ahe4,ahe5}.
If the spatial variance of the magnetization dynamics is slow enough, Eq.~(\ref{magnetic}) is proportional to the solid angle subtended by the magnetization triangle, $\textrm{\boldmath $m$}_1\cdot\left(\textrm{\boldmath $m$}_2\times\textrm{\boldmath $m$}_3\right)$, which is called scalar chirality. 
$\textrm{\boldmath $m$}_i$ represents the magnetization at site $i$.

\begin{figure}[t]
  \centerline{\includegraphics[width=80mm]{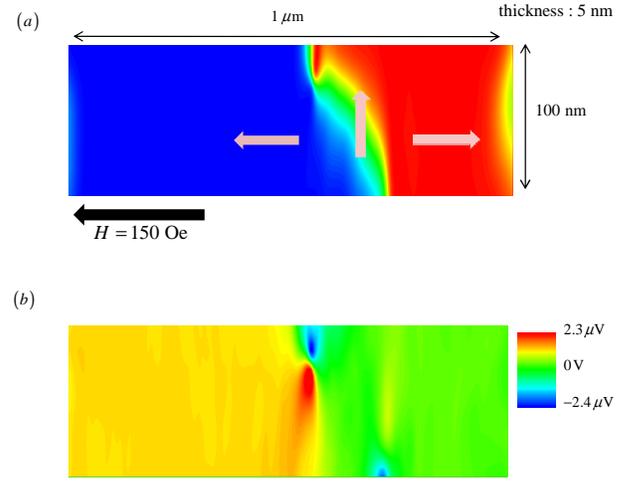}}
  \caption{
   Numerical calculations of the spin-motive force with the domain wall motion in a Ni$_{81}$Fe$_{19}$ thin wire.
   (a) The magnetization configuration at a certain time.
   Pink arrows indicate the local magnetizations.
   An external magnetic field $H=150$ Oe is applied along the wire direction.
   The material parameters are : $P=0.6$, $M_S=1$ T, the exchange stiffness is $A=1.3\times 10^{-11}$ J/m, the Gilbert damping constant $\alpha=0.01$ and $\gamma=1.76\times 10^{11}$ Hz/T.
   We used OOMMF code\cite{oommf} and the unit cell size is $4\times 4\times 5$ nm$^3$.
   (b) The electric potential distribution corresponding to the magnetization dynamics shown in (a).
 }
 \label{fig:fig3}
\end{figure}

Here we summarize how to calculate the spin-motive force.
As an example, we demonstrate a numerical calculation of the spin-motive force due to the field induced domain wall motion in a ferromagnetic thin wire.
First, we have to know the magnetization dynamics $\textrm{\boldmath $m$}$ by solving the LL equation (\ref{ll}) in a given condition [Fig.~\ref{fig:fig3}(a)].
Then, the spin dependent force acting upon an electron in each time and space can be calculated by Eq.~(\ref{force3}).
The spin-motive force is obtained by the spatial difference of the electric potential $V$ [Fig.~\ref{fig:fig3}(b)], which is given by the Poisson equation $\frac{P}{-e}\nabla\cdot\left\langle f\right\rangle_{\uparrow}=\Delta V$ \cite{ohe}.
Here, $e$ and $P$ are the elementary charge and the spin polarization of the conduction electrons, respectively.

%
%

\section*{summary}

In summary, we have disccussed the electron charge dynamics in ferromagnets by using the equation-of-motion approach, and derived a general formula of the spin dependent force acting upon electrons due to the exchange interaction.
It has been revealed that the origin of the spin-motive force is the misalignment between the electron spin and the local magnetization.

%
%

\section*{acknowledgments}

The authors are grateful to E. Saitoh, K. Sasage, M. Mori, S. Takahashi, S. Hikino, and M. Matsuo for valuable discussions.
This work was supported by Grant-in-Aid for Scientific Research from MEXT, Japan, 
and the Next Generation Supercomputer Project, 
Nanoscience Program from MEXT, Japan.

%
%

\end{document}